# Relaxation mechanism driven by spin angular momentum absorption throughout antiferromagnetic phase transition in NiFe surface oxides

L. Frangou,[1] G. Forestier,[1] S. Auffret,[1] S. Gambarelli,[2] and V. Baltz[1,*]

[1] SPINTEC, Univ. Grenoble Alpes / CNRS / INAC-CEA, F-38000 Grenoble, France
[2] SYMMES, Univ. Grenoble Alpes / INAC-CEA, F-38000 Grenoble, France
[*] vincent.baltz@cea.fr

**Abstract**

We report an alternative mechanism for the physical origin of the temperature-dependent ferromagnetic relaxation of Permalloy (NiFe) thin films. Through spin-pumping experiments, we demonstrate that the peak in the temperature-dependence of NiFe damping can be understood in terms of enhanced spin angular momentum absorption at the magnetic phase transition in antiferromagnetic surface-oxidized layers. These results suggest new avenues for the investigation of an incompletely-understood phenomenon in physics.

In physical systems, damping characterizes the losses associated with out-of-equilibrium vibration dynamics [1,2]. In the field of spintronics, which relies on the spin-dependent transport properties of matter [3–5], magnetic damping is one of the key parameters as it regulates oscillations and switches in magnetization direction in any magnetic material [6,7]. Magnetic damping plays this role with all kinds of stimuli, whether the dynamics of magnetization is excited through an electromagnetic wave [8], an electrical current [9], or a spin current [10]. Damping in typical ferromagnetic materials has been thoroughly experimentally characterized through measurements of ferromagnetic resonance spectra and determination of their linewidths [11]. Several mechanisms have been proposed to explain these findings [11]. However, the basic mechanisms behind some magnetic relaxation behavior remain unclear even in common magnetic materials. For example CoFe alloys were recently theoretically predicted [12] and experimentally demonstrated [13] to display ultra-low damping which was previously believed to be unachievable in metallic ferromagnets. In this article, inspired by recent theoretical and experimental findings on spin-pumping [14–19], we chose to investigate Permalloy (NiFe) in an attempt to determine the incompletely-understood origin of their non-monotonous temperature-dependence of ferromagnetic damping [20–25]. More specifically, typical 3*d* transition metals (Co, Ni, Fe) and associated alloys (including NiFe) frequently show a minimum in their temperature-dependence of damping [26,27]. It is now accepted that a conductivity-like term related to intraband scattering dominates at low temperatures, whereas a resistivity-like term due to interband scattering takes over at higher temperatures [27]. Sometimes for NiFe, a contrasting pronounced maximum has been reported in the temperature-dependent damping superimposed with a monotonous decreasing baseline [20]. This finding, and the reasons for it, remain controversial and are still being discussed. It has been suggested that the temperature-dependent reorientation of NiFe surface spins from in-plane to out-of-plane could account for the maximum damping observed [22–24].

However, recent findings show that the spin reorientation may occur at a much lower temperature than the maximum damping [28]. An alternative mechanism was also proposed, involving slow relaxation on paramagnetic impurities present in or adjacent to the oscillating ferromagnetic material [20,21,29,30]. In this process, the oscillations in the magnetization of the ferromagnet modulate how the energy splits between impurity levels. Subsequent relaxation of the impurities influences ferromagnetic damping. In fact, if not protected from oxidation due to exposure to air, a few monolayers of the NiFe layer will naturally oxidize to form a passivating oxide layer (NiFeOx). This layer contains a complex mixture of NiO and FeO antiferromagnetic alloys with variable stoichiometry gradients [25]. In this context, the potential influence of relaxation of interface paramagnetic impurities in bilayers where a ferromagnet is exchange-biased to an antiferromagnet was considered in several studies [31–34]. The results of these studies led to divergent mechanisms for the temperature-dependence of the relaxation rate for impurities [31–34]. Beyond paramagnetic impurities or exchange-bias interactions, the presence of NiFe antiferromagnetic surface oxides raises the question of how spin angular momentum is absorbed by the antiferromagnetic layer itself [35,36]. In this process, transfer/sink and propagation of spin angular momentum involves magnons from the oscillating ferromagnet feeding into the entire antiferromagnet, due to magnetic coupling [37,38]. The end result is an overall enhancement of the intrinsic damping of the ferromagnet [35,36]. In addition, near the phase transition for the magnetic order of the antiferromagnetic layer, i.e., around its Néel temperature, the magnetic fluctuations were shown to lead to a maximum spin-pumping efficiency [16–18]. The origin of this phenomenon was corroborated by calorimetry [16,39] and neutron diffraction measurements [18].

In this work, we investigated whether enhanced spin angular momentum absorption at the magnetic phase transition of surface-oxidized layers could be an alternative mechanism explaining the temperature-dependent ferromagnetic relaxation of NiFe. We examined temperature-dependent ferromagnetic relaxation in NiFe thin films, and how it was affected by oxidation of the NiFe layer and the number of oxide layers surrounding the NiFe (two, one or none). Spin-pumping experiments were performed at various temperatures on two series of samples. The first series consisted of Si/SiO$_2$(500)/NiFe(8)-Ox (short name: SiO$_2$ / NiFe-Ox), Si/SiO$_2$(500)/NiFe(8)/Cu(3)/Al(2)-Ox (short name: SiO$_2$ / NiFe / Cu), Si/SiO$_2$(500)/Cu(6)/NiFe(8)-Ox (short name: Cu / NiFe-Ox) and Si/SiO$_2$(500)/Cu(6)/NiFe(8)/Cu(3)/Al(2)-Ox (short name: Cu / NiFe / Cu) multilayers. All thicknesses are given in nanometers and -Ox stands for oxidation in air. The second series consisted in Si/SiO$_2$(500)/NiFe($t_{NiFe}$)-Ox/NiFe(8)-Ox multilayers, where $t_{NiFe}$ is the thicknesses of the bottom NiFe layer - 0.5, 1, or 1.5 nm. Stacks were deposited on thermally oxidized silicon substrates [Si/SiO$_2$(500)] at room temperature by dc-magnetron sputtering. The NiFe layer was deposited from a Permalloy target [Ni$_{81}$Fe$_{19}$ (at. %)]. An Al(2) cap was added to block oxidization by air in some samples, this cap forms a protective passivating AlOx film. Transmission electron microscopy (TEM) analysis (Fig. 1(a)) was used to view oxidation of the NiFe layer in the SiO$_2$ / NiFe-Ox stack. Results of these investigations indicated that the thickness of the NiFeOx surface oxide due to NiFe oxidation in air is approximately 1.6 ± 0.2 nm. This value is in line with data from the literature, where passivating surface oxides were reported to measure nanometers thick [25]. Representative results from energy dispersive x-ray spectroscopy (EDX) measurements are illustrated in Fig. 1(b). These data confirm the presence of a surface-oxidized layer and reveal the presence of another oxidized layer at the interface between the SiO$_2$ and NiFe layers. This lower oxide layer was not visible in the TEM image due to a lack of contrast with the SiO$_2$ underlayer. The presence and thickness (around 0.3 ± 0.2 nm) of this bottom oxide layer was determined from the horizontal shift in the oxygen and silicon traces in Fig. 1(b). From the data shown in Fig. 1(b) we also calculated that in the SiO$_2$

/ NiFe-Ox sample the Ni and Fe atoms extend over a total thickness of around 8.1 ± 0.2 nm. Complementary EDX measurements of a $SiO_2$ / NiFe / Cu sample, where the NiFe layer was not air-oxidized, indicated that the Ni and Fe atoms also extend over a total thickness of around 8 ± 0.3 nm, suggesting a negligible expansion of the lattice parameter for the oxide layer in the $SiO_2$ / NiFe-Ox samples.

We next investigated the magnetic nature of the surface-oxidized layers by measuring hysteresis loops at various temperatures using a magnetometer (Fig. 2(a)). These results show a loop shift ($H_E$) along the axis of the magnetic field, demonstrating magnetic exchange-bias interactions [40,41] between the NiFe ferromagnetic layer and the NiFeOx surface-oxidized layer. These data thus confirm the antiferromagnetic nature of the top surface-oxidized layer. The data presented in Fig. 2(b) further indicated that $H_E$ decreases as the temperature rises. The ferromagnetic/antiferromagnetic blocking temperature ($T_B$) can be extracted from $H_E$ vs. T by determining the temperature at which $H_E$ vanishes [40,41]. For the NiFe/NiFeOx(1.6) bilayer, $T_B$ was found to be about 15 K (see data for the Cu / NiFe-Ox and $SiO_2$ / NiFe-Ox samples). $T_B$ is expected to be much smaller than the critical temperature ($T_{crit}$) for the antiferromagnetic to paramagnetic transition [40,41]. This relationship can be explained as $T_B$ relates to the interfacial exchange interactions between the ferromagnet and the antiferromagnet, whereas $T_{crit}$ relates to the exchange stiffness between all antiferromagnetic moments. For the NiFeOx(0.3)/NiFe bilayer (see data for the $SiO_2$ / NiFe / Cu sample) $T_B$ is sub-K and could not be measured based on the data shown in Fig. 2(b) due to the fact that the lower NiFeOx oxide layer is very thin and displays a reduced $T_{crit}$. Note that for the NiFeOx(0.3) ultra-thin layer, $T_{crit}$ probably describes a frozen to liquid spin transition. Results confirming the reduced value of $T_{crit}$ will be discussed further.

Spin-pumping experiments [Fig. 3(a)] and related series of ferromagnetic resonance spectra were recorded for temperatures (T) ranging between 20 and 300 K, using a continuous-wave electron paramagnetic resonance spectrometer operating at 9.6 GHz fitted with a cavity. For each temperature the Gilbert damping ($\alpha$) was determined by fitting the NiFe resonance spectrum to a Lorenzian. The value of $\alpha$ was extracted from: $\alpha(T) = \left[\Delta H_{pp}(T) - \Delta H_0(300K)\right]\sqrt{3}|\gamma|/2\omega$, where $\Delta H_{pp}$ is the peak-to-peak linewidth for the spectrum, $\gamma$ is the gyromagnetic ratio, and $\omega$ is the angular frequency [42]. $\Delta H_0$ relates to spatial variations in the magnetic properties. This parameter was determined from standard $\Delta H_{pp}$ vs. $\omega/2\pi$ plots using a separate, broadband coplanar waveguide operating at room temperature for frequencies ranging between 2 and 24 GHz [42]. Figure 3(b) shows $\alpha$ plotted against temperature, the pronounced maximum at T = 70 K corresponds to the air-oxidized NiFe layer (see data for the $SiO_2$ / NiFe-Ox and Cu / NiFe-Ox samples). Its amplitude is 3-fold greater than the amplitude measured at 300 K. A less pronounced contribution is visible at lower temperatures in samples where the NiFe oxidized in contact with the $SiO_2$ layer (see data for the $SiO_2$ / NiFe / Cu sample). When the NiFe layer was isolated from oxygen atoms in the Cu / NiFe / Cu sample no such maximum was observed. Since the oxidized layers are magnetic, the NiFe damping is the sum of local intrinsic damping ($\alpha^0$) and additional non-local damping ($\alpha^{p,i}$) associated with the surface/interface oxide(s) acting as a spin-sink for angular momentum. The temperature-dependence of $\alpha$ can be expressed as: $\alpha(T) = \alpha^0(T) + \sum_i \alpha^{p,i}(T)$ [14,35,36], where $i$ accounts for the uppermost and/or lowermost NiFeOx spin absorber. Data obtained with the Cu / NiFe / Cu sample (no spin absorber) give the temperature-dependence of the local intrinsic NiFe Gilbert damping [$\alpha_{Cu/NiFe/Cu}(T) = \alpha^0(T)$] with a detectable conductivity- to resistivity-like progression [26,27]. From Fig. 1(b), we can thus conclude that the temperature-dependence of $\alpha^0$ can be neglected, but that $\alpha^{p,i}$ is highly temperature-dependent. The non-local damping is given as presented in [14] by: $\alpha^{p,i}(T) = \dfrac{2|\gamma|J_{sd}^2 S_0 N_{int}}{\hbar N_{SA}^i N_{SI} M_{S,NiFe} t_{NiFe}} \sum_k \dfrac{1}{\Omega_{rf}} \operatorname{Im} \chi_k^{R,i}(\Omega_{rf}, T)$

, where $S_0$ is the norm of the spin operator, $N_{SI}$ is the number of lattice sites in the NiFe spin injector (SI), $N_{int}$ is the number of spins localized at the interface, $N_{SA}^i$ is the number of lattice sites in the spin absorber (SA) $i$, $J_{sd}$ is the s-d exchange interaction at the SI/SA interface, $k$ is the wave vector, $\Omega_{rf}$ is the NiFe angular frequency at resonance, and $t_{NiFe}$ is the thickness of the NiFe layer. The temperature-dependent dynamic spin susceptibility of the NiFeOx oxide is represented by $\chi_k^{R,i}(\Omega_{rf}, T)$. The spin susceptibility of antiferromagnetic materials is known to display a maximum around the critical temperature for the magnetic phase transition. This transition results in enhanced spin angular momentum absorption and translates into maximal NiFe damping, as observed in Fig. 3(b). From the Cu / NiFe-Ox data, where $\alpha_{Cu/NiFe-Ox}(T) = \alpha^0(T) + \alpha^{p,NiFeOx(1.6)}(T)$, we deduced the Néel temperature for the magnetic phase transition of the top 1.6-nm-thick NiFeOx oxide, at approximately 70 K. From the SiO$_2$ / NiFe / Cu data in Fig. 3(b), where $\alpha_{SiO_2/NiFe/Cu}(T) = \alpha^0(T) + \alpha^{p,NiFeOx(0.3)}(T)$, we concluded that the critical temperature for the phase transition of the lowermost 0.3-nm-thick NiFeOx oxide, which formed at the interface between the NiFe and SiO$_2$ layers, is less than 20 K. We infer that this temperature is actually well below 20 K, and probably sub-K since the amplitude of the damping peak for the 0.3-nm-thick NiFeOx oxide is expected to be 5 times (1.6/0.3) larger than that of the 1.6-nm-thick oxide. The reason for this difference is that $\alpha^p$ is inversely proportional to the number of lattice sites in the spin absorber ($N_{SA}$). Finally, the data for the SiO$_2$ / NiFe-Ox sample relate to $\alpha_{SiO_2/NiFe-Ox}(T) = \alpha^0(T) + \alpha^{p,NiFeOx(0.3)}(T) + \alpha^{p,NiFeOx(1.6)}(T)$. As shown in Fig. 3(b), we verified that $\alpha_{SiO_2/NiFe-Ox} - \alpha_{SiO_2/NiFe/Cu} + \alpha_{Cu/NiFe/Cu} = \alpha_{Cu/NiFe-Ox}$.

We further investigated how $T_{crit}$ depends on the thickness of the oxidized layer. Figure 4(a) shows α plotted against temperature for Si/SiO$_2$(500)/NiFe($t_{NiFe}$)-Ox/NiFe(8)-Ox multilayers with $t_{NiFe}$ = 0.5, 1 and 1.5 nm. Based on the results presented above, the lowermost

NiFe layer is expected to be fully oxidized. The samples therefore consisted of a NiFe layer sandwiched between two NiFeOx spin angular momentum absorbers. The data shown in Fig. 4(a) indicate two peaks in α for samples containing the 0.5- and 1-nm thick lowermost NiFeOx layers. The peak at around 70 K corresponds to the magnetic phase transition of the NiFeOx layer on top. The peak at the lower temperature corresponds to the magnetic phase transition of the bottom NiFeOx layer. From Fig. 4(a), we can see that the contribution of the phase transition of the lower layer shifts towards higher temperatures as its thickness increases. With samples containing the 1.5-nm thick lower NiFe oxidized layer, the peaks corresponding to the magnetic phase transition of the top and bottom NiFeOx layers overlapped. The peak's amplitude is close to twice the amplitude of the peak for the Cu / NiFe - Ox sample (which only contains the top 1.6 nm NiFeOx layer). This observation indicates that the top and bottom layers absorb similar amounts of spin current on both sides and share a similar $T_{crit}$. Figure 4(b) illustrates how the critical temperature for the NiFeOx layer is directly proportional to its thickness. This linear relationship is in line with theories on finite size scaling of magnetic phase transitions [43,44] whereby $T_{crit}(t_{NiFeOx})=T_N(bulk)(t_{NiFeOx}-d)/(2n_0)$, with $T_N(bulk)$ as the bulk Néel temperature of the NiFeOx layer, $t_{NiFeOx}$ as its thickness, $d$ as its lattice parameter, and $n_0$ as its phenomenological inter-spin correlation length. Our data cannot be readily fitted to the model because the nature of the NiFeOx layer is complex, composed of a mixture of different phases including NiO and CoO alloys (approximately proportional to the initial Ni-to-Fe 20/80 atomic ratio) and thickness gradients in the oxidation rate [25]. The red line in Fig. 4(b) is a fit for the $Ni_{81}Fe_{19}Ox$ layer based on considering it as a $(NiO)_{81}(FeO)_{19}$ alloy. We used $T_N(bulk)=0.81T_{N,NiO}(bulk)+0.19T_{N,FeO}(bulk)$ for fitting, with $T_{N,NiO}(bulk) = 520$ K, $T_{N,FeO}(bulk) = 200$ K [40], $d=0.81d_{NiO}+0.19d_{FeO}$, $d_{NiO} = 0.417$ nm, and $d_{FeO} = 0.433$ nm. The fit agrees with our data to a satisfactory extent, and returned $n_0 = 4.4$ nm (approximately ten monolayers), which is typical for ordered magnetic films [25].

In conclusion, the main contribution of this paper is the experimental evidence it presents supporting an alternative mechanism explaining the incompletely-understood physical origin of the temperature-dependent ferromagnetic relaxation of Permalloy. Our results demonstrated that the peak in temperature-dependence of Permalloy damping can be understood in terms of enhanced absorption of spin angular momentum at the antiferromagnetic to paramagnetic phase transition of its surface-oxidized layer. These findings open perspectives for further investigations since a multitude of magnetic materials form antiferromagnetic spin absorbers upon oxidation.

**Acknowledgments**

We acknowledge financial support from the French National Agency for Research [ANR JCJC ASTRONICS, Grant Number ANR-15-CE24-0015-01]. We also thank M. Gallagher-Gambarelli for providing advice on English usage.

**Figure captions**

Fig. 1. (color online) (a) Transmission electron microscopy image (TEM) and (b) energy-dispersive x-ray spectroscopy (EDX) data for a Si/SiO$_2$(500)/NiFe(8)-Ox (nm) sample. Samples were capped with Pt in preparation for the TEM experiment.

Fig. 2. (color online) (a) Representative magnetization (M) *vs.* field (H) hysteresis loops at different temperatures for a Si/SiO$_2$(500)/Cu(6)/NiFe(8)-Ox (nm) sample. (b) Temperature (T)-dependence of the hysteresis loop shift (H$_E$).

Fig. 3. (color online) (a) Diagrammatic representation of the spin-pumping experiment. (b) Temperature (T)-dependence of the NiFe layer Gilbert damping ($\alpha$). The NiFe layer is oxidized in air or not, and surrounded by two, one or no oxide layer.

Fig. 4. (color online) (a) Temperature (T)-dependence of the Gilbert damping ($\alpha$) of the NiFe(8) layer on temperature (T) in Si/SiO$_2$(500)/NiFe(t$_{NiFe}$)-Ox/NiFe(8)-Ox multilayers. (b) Thickness-dependence of the critical temperature (T$_{crit}$) for the magnetic phase transition of the oxidized NiFe layer. Open circles represent data deduced from Fig. 4(a), they are plotted against the initial NiFe thickness (t$_{NiFe}$). Full squares represent data deduced from Fig. 3(b) and the corresponding text; they are plotted against the NiFeOx thickness determined from TEM and EDX experiments (see Fig. 1). Line fitting was based on the equation presented by Zhang et al. [43] in the thin-layer regime for a (NiO)$_{81}$(FeO)$_{19}$ alloy.

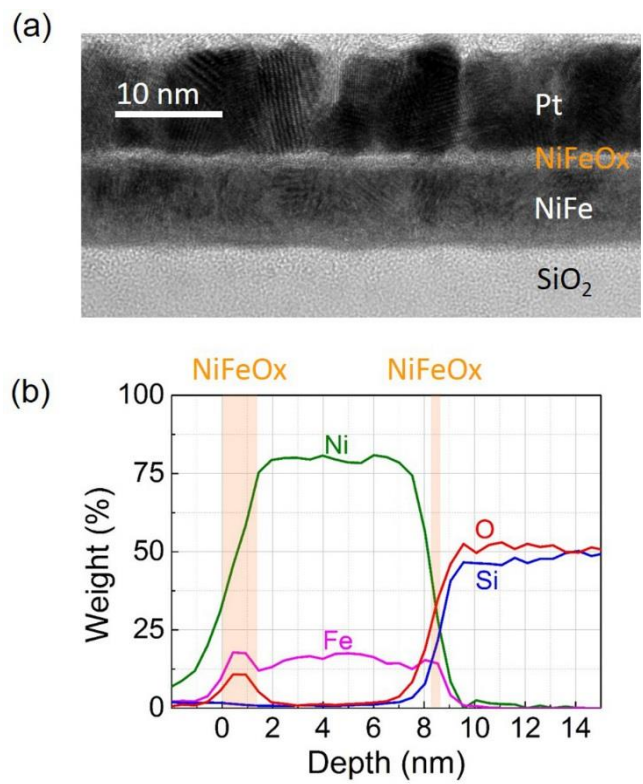

L. Frangou et al Fig. 1

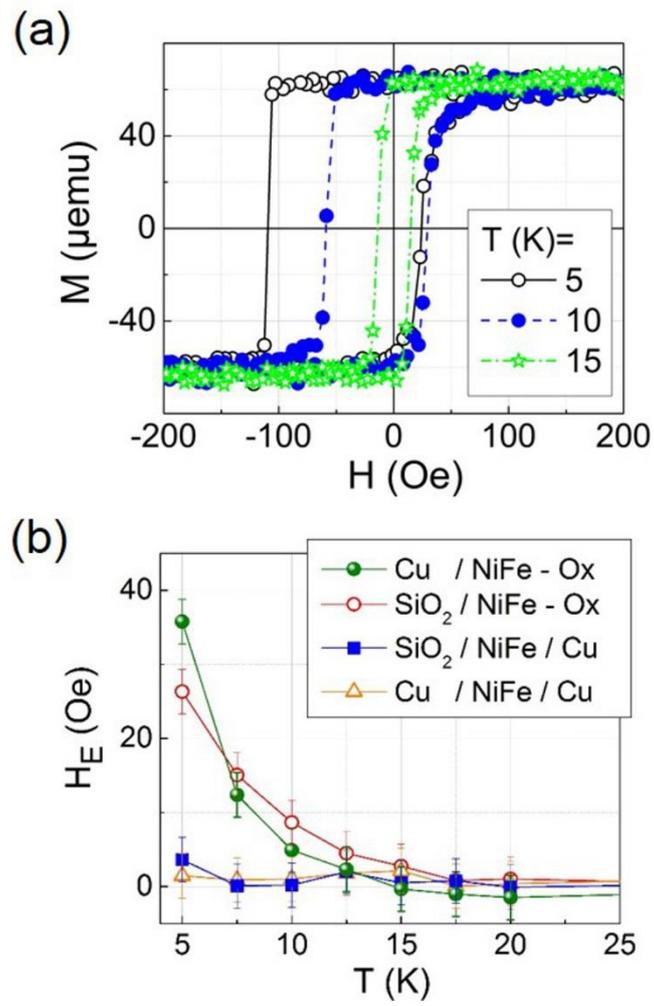

L. Frangou et al                                                                                                                  Fig. 2

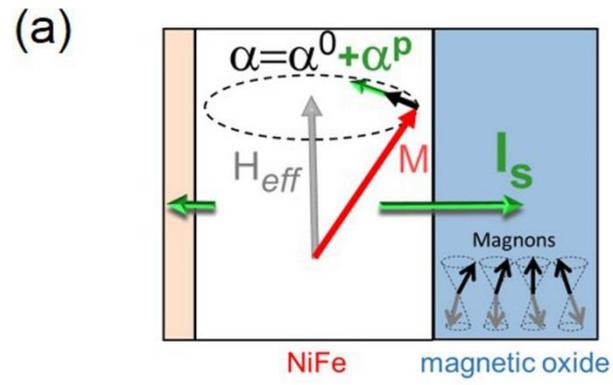

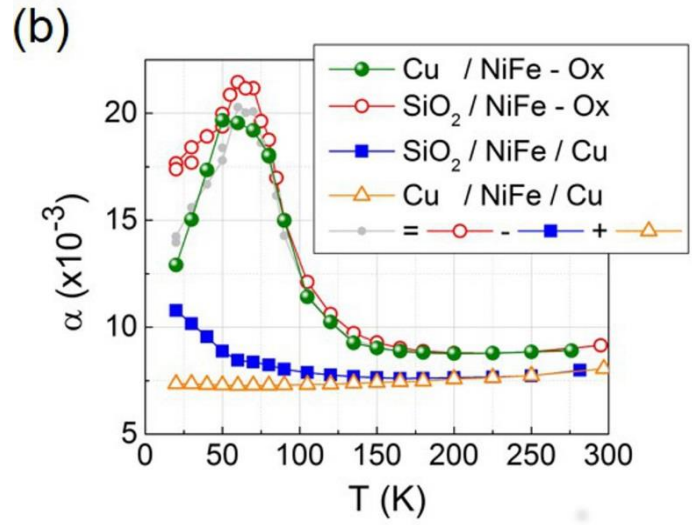

L. Frangou et al                                           Fig. 3

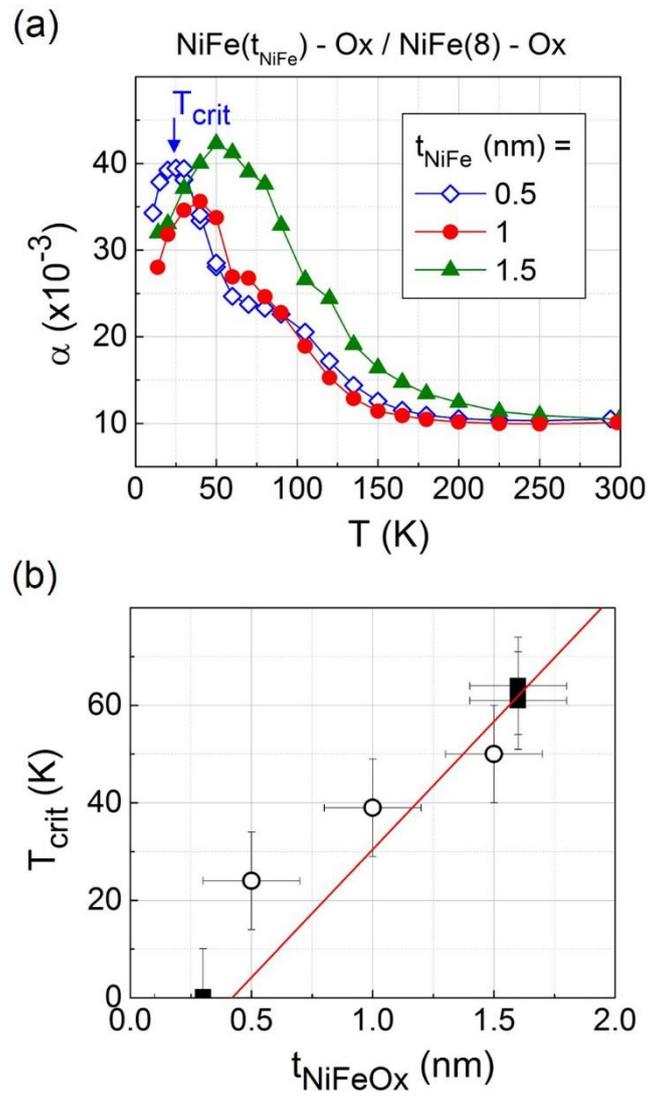

L. Frangou et al					Fig. 4